\documentclass[11pt,a4paper]{article}
\usepackage{amsmath,amssymb}
\usepackage{epsfig,graphicx}
\usepackage{epic,eepic}
\usepackage{subfigure}
\usepackage{rotating}
\usepackage{bm}
\usepackage{axodraw4j}
\usepackage{color}
\usepackage{hyperref}
\usepackage{mathabx}

\newlength{\wth}
\def\CH{{\cal H}}

\def\Dbarslash{\,\,{\raise.15ex\hbox{/}\mkern-12mu {\bar\D}}}
\def\Dslash{\,\,{\raise.15ex\hbox{/}\mkern-12mu \D}}
\def\delslash{\,\,{\raise.15ex\hbox{/}\mkern-9mu \partial}}
\def\delbarslash{\,\,{\raise.15ex\hbox{/}\mkern-9mu {\bar\partial}}}

 \def\L{\Lambda}

\newcommand{\gev}{\mbox{\,GeV\,}}
\newcommand{\tev}{\mbox{\,TeV\,}}

\def\D{{\cal D}}
\def\Dbarslash{\,\,{\raise.15ex\hbox{/}\mkern-12mu {\bar\D}}}
\def\delslash{\,\,{\raise.15ex\hbox{/}\mkern-9mu \partial}}
\def\Dslash{\,\,{\raise.15ex\hbox{/}\mkern-12mu \D}}

\def\={\, =\, }
\def\+{\, +\, }
\def\-{\, -\, }

\newcommand{\be}{\begin{equation}}
\newcommand{\ee}{\end{equation}}
\def\bea{\begin{eqnarray}}
\def\eea{\end{eqnarray}}

\definecolor{saabeer}{rgb}{0,1,0}

\definecolor{durbeer}{rgb}{1,0,0}
\newcommand{\JJ}[1]{{\color{durbeer} #1}}
\definecolor{durbeer2}{rgb}{0,0,0.7}

\begin{document}
\date{\mbox{ }}
\title{{\normalsize  IPPP/10/57; DCPT/10/114; CERN-PH-TH/2010-200; DAMTP-2010-60\hfill\mbox{}\hfill\mbox{}}\\
\vspace{2.5cm} \LARGE{\textbf{Pure General Gauge Mediation \\ for Early LHC Searches}}}
\author{Steven Abel$^{a,b}$, Matthew J. Dolan$^c$, Joerg Jaeckel$^a$ and  Valentin V. Khoze$^a$\\[2ex]
\small{$^a$\em Institute for Particle Physics Phenomenology,}\\
\small{\em Durham University, Durham DH1 3LE, United Kingdom}\\[2ex]
\small{$^b$\em TH Division, Physics Department, CERN,}\\
\small{\em CH-1211 Geneva 23, Switzerland}\\[2ex]
\small{$^c$ \em DAMTP, CMS, University of Cambridge, Wilberforce Road,}\\
\small{\em Cambridge, CB3 0WA, United Kingdom}}  
\date{}
\maketitle

\vspace{3ex}

\begin{abstract}
\noindent We present benchmark points for Pure General Gauge Mediation (GGM) models specifically optimised
for early LHC searches. The pure GGM set-up is as defined in our previous paper
\href{http://arXiv.org/abs/0910.2674}{arXiv:0910.2674}:
namely we adopt the minimal set-up in which $B_\mu$ is generated only through gauge interactions, and as a
result $\tan\beta$ is a prediction rather than an input. The only input parameters are messenger masses
together with two independent scales which generate gaugino and scalar masses.
The parameter space favoured by current experimental data
includes an interesting region with light gluinos ($m_{\tilde{g}} \lesssim 500\gev$ and
relatively heavy squarks) and a bino-like NLSP where early discovery is likely.
We also find interesting regions of parameter space where the NLSP is a stau or stau/neutralino co-NLSP
and the squark masses are relatively low.
We present  benchmark points which are typical for each of these three regions,
and give the spectrum, branching ratios, and also the overall
$2\rightarrow 2$ cross-sections. This data in SLHA format can be found at
\href{http://www.ippp.dur.ac.uk/~SUSY}{\bf http://www.ippp.dur.ac.uk/$\sim$SUSY}.

\end{abstract}

\newpage


\section{Introduction}

\noindent Ref.~\cite{Meade:2008wd} introduced a novel framework suitable for discussing and analysing
general models of gauge mediation in a model-independent way. The so-called General Gauge Mediation
(GGM) paradigm~\cite{Meade:2008wd} is defined by the requirement that the Minimal Supersymmetric Standard
Model (MSSM) becomes decoupled from the hidden SUSY-breaking sector in the limit where the three MSSM
gauge couplings $\alpha_{i=1,2,3}$ are set to zero. Since no other parameters participate
in the coupling of the two sectors, we call this strict interpretation of gauge mediation
`general pure gauge mediation' or {\em pure} GGM. This framework is broad enough to include everything from
weakly coupled models with explicit messengers to strongly coupled theories with direct mediation.

Preliminary investigations of the phenomenology of GGM have been made
in Refs.~\cite{Carpenter:2008he,Rajaraman:2009ga,Abel:2009ve}. In particular
Ref.~\cite{Abel:2009ve} concentrated on the pure GGM scenario which we shall be
adopting here. This is in a sense the most minimal assumption because it obviates the
need for an additional sector just to generate the bilinear $B_\mu$ parameter for the
higgses. To summarise the approach, in addition to the supersymmetric interaction,
\be
\label{mudef}
{\cal L}_{eff}\supset\int d^2 \theta \,\, \mu \,\CH_u\CH_d \ ,
\ee
the Higgs-sector effective Lagrangian also includes
soft supersymmetry-breaking terms. All of the latter must be
generated by the SUSY-breaking sector,
since there would be little merit in
a model of dynamical SUSY-breaking which generates only a
subset of the SUSY-breaking terms in the effective SM Lagrangian.
There are quadratic terms
\be
\label{quaddef}
m_u^2 |H_u|^2 + m_d^2 |H_d|^2 +(B_\mu H_uH_d + c.c.)~,
\ee
as well as cubic $a$-terms
\be
\label{Atermsdef}
a_u^{ij} H_u Q^i \bar u^j + a_d^{ij} H_d Q^i \bar d^j + a_L^{ij} H_d L^i \bar E^j~,
\ee
in the MSSM. As is well-known, a phenomenologically acceptable electroweak
symmetry breaking in the supersymmetric SM occurs
if $\mu^2$ and the soft masses in \eqref{quaddef} at the low scale
(i.e. the electroweak scale)
are of the same order, $\mu^2 \sim B_\mu \sim m_{soft}^2\sim M^2_{W}$.

In pure GGM we have no direct couplings
of the SUSY-breaking sector to the Higgs sector, and therefore must have
$B_{\mu}\approx 0$ at the
messenger scale. From this starting point, i.e. taking $B_{\mu}\approx 0$ at the
high scale $M_{mess}$, a small but viable value of $B_\mu$ is
generated radiatively at the electroweak scale \cite{Rattazzi:1996fb,Babu:1996jf}.
Electroweak symmetry breaking then determines the
values of $\tan\beta$ and $\mu$.
Since $B_\mu$ is small, $\tan\beta$ is generally
large (between $20$ and $70$).
This setup where $B_{\mu}\approx 0$ is an input and $\tan \beta$ is an output \cite{Abel:2007nr,Abel:2009ve}
is in contrast to the common approach where $\tan\beta$ is taken as an arbitrary
input and $B_{\mu}$ at the high scale is obtained from it.

The main free parameters are the gaugino and scalar masses as well as
the messenger scale. For simplicity we restrict ourselves to a single effective
scale $\L_G$ for the gaugino masses and a single scale $\L_S$ for the
scalars\footnote{Of course in each specific GGM model, the parameters $\L_G$ and $\L_S$ determining gaugino
and scalar masses at the messenger scale are computed and expressed in terms of the scales of the SUSY-breaking sector,
and details of the messenger fields. As such, $\L_G$ and $\L_S$ (together with $M_{mess}$) characterise a point in the pure GGM
parameter space and can be treated as input parameters. }.
Thus at the messenger scale $M_{mess}$ the soft supersymmetry breaking gaugino masses are
\begin{equation}
\label{gauginosoft}
M_{\tilde{\lambda}_i}(M_{mess}) =\, k_i \,\frac{\alpha_i(M_{mess})}{4\pi}\,\Lambda_G
\end{equation}
where $k_i = (5/3,1,1)$, $k_i\alpha_i$ (no sum)
are equal at the GUT scale and $\alpha_i$ are the gauge coupling constants.
The scalar mass squareds are
\begin{equation}
\label{scalarsoft}
m_{\tilde{f}}^2 (M_{mess}) =\, 2 \sum_{i=1}^3 C_i k_i \,\frac{\alpha_i^2(M_{mess})}{(4\pi)^2}\, \Lambda_S^2
\end{equation}
where the $C_i$ are the quadratic Casimir operators of the gauge groups.
Ordinary gauge mediation scenarios (see Ref.~\cite{Giudice:1998bp} for a review) live on the restricted
parameter space $\L_G\simeq\L_S$. We have implemented these boundary conditions in a modified version of \texttt{Softsusy}~\cite{Allanach:2001kg}, which takes $_{\mu}$ as an input and predicts $\tan\beta$ using the electroweak symmetry breaking conditions.

\begin{figure}
\begin{center}
\begin{picture}(190,180)
\includegraphics[viewport= 160 70 380 380, width=6cm]{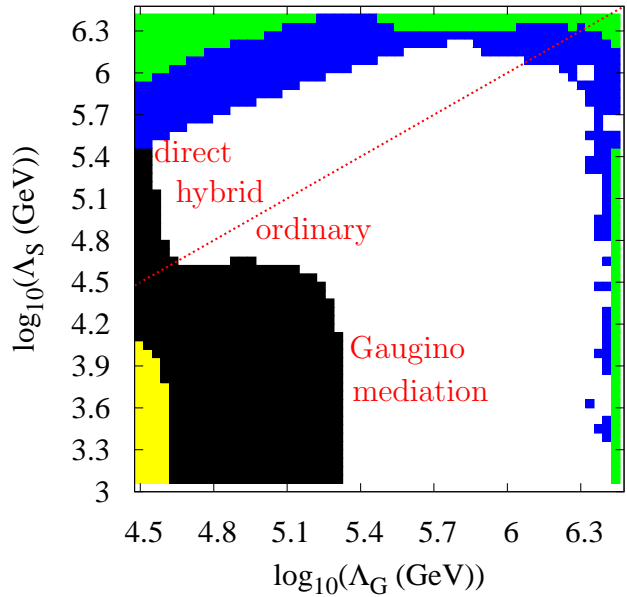}
\Text(-155,135)[c]{\scalebox{1}[1]{\large \JJ{direct}}}
\Text(-145,119)[c]{\scalebox{1}[1]{\large \JJ{hybrid}}}
\Text(-110,105)[c]{\scalebox{1}[1]{\large \JJ{ordinary}}}
\Text(-75,60)[c]{\scalebox{1}[1]{\large \JJ{Gaugino}}}
\Text(-70,45)[c]{\scalebox{1}[1]{\large \JJ{mediation}}}
\end{picture}
\vspace*{1cm}
\end{center}
\begin{center}
\caption{
The underlying mediation physics corresponding to 
different regions of the $\Lambda_G$,~$\Lambda_S$ parameter space. 
In the extreme $\Lambda_G\ll \Lambda_S$ region we 
have direct gauge mediation with no lower lying tree-level minima. 
Outside this region lies the hybrid region with lower lying minima being 
brought in from infinity. The red dotted line indicates the ordinary gauge 
mediation line where $\Lambda_G = \Lambda_S$, which can be reproduced 
in metastable set-ups with high messenger scales such as those in Ref.~\cite{Murayama:2006yf}. 
Below the ordinary gauge mediation line we find the ``many effective messenger'' 
$\Lambda_G \gg \Lambda_S$  region,
which is where some mechanism screens the contributions to the scalar masses. We also show the 
allowed region for intermediate messenger scales,  $ M_{Mess} = 10^{10}$~GeV with the 
dominant constraints excluding various areas indicated as follows: 
yellow (pale grey) means the point is excluded by the presence of tachyons in the spectrum, while the black region falls foul of the direct search limits. In the blue (dark grey) region SoftSUSY has not converged and in the green (light grey) region a coupling reaches a Landau pole during RG evolution.}
\label{phenoland}
\end{center}
\end{figure}

Outside the confines of \emph{ordinary} gauge mediation the
parameter space is populated by many models that
predict different values of the ratio of gaugino to scalar masses, $\L_G/\L_S$. In
models with explicit messengers one expects
this ratio {to be} close
to one, while for direct mediation models the gaugino masses are
often suppressed relative to the scalar
masses~\cite{Izawa:1997gs,Kitano:2006xg,Csaki:2006wi,Abel:2007jx,Abel:2007nr,Abel:2008gv}. 
Ref.~\cite{Komargodski:2009jf} provided a general argument that linked the
gaugino mass to the existence of lower lying minima at tree-level. 
Indeed hybrid models can easily be constructed which interpolate
between these two cases by bringing lower lying minima in from infinity~\cite{Abel:2009ze}. It is also possible to
achieve values $\L_G/\L_S>1$ by increasing the ``effective number of
messengers'' in the context of extraordinary gauge mediation
models~\cite{Cheung:2007es}.
Naively this ``gaugino mediation'' region of parameter space corresponds to strong coupling, 
but explicit and calculable models space are possible in the context of extra dimensional  
models \cite{Mirabelli:1997aj,Kaplan:1999ac,Chacko:1999mi,Csaki:2001em,McGarrie:2010kh,McGarrie:2010qr} or 
electric/magnetic duality~\cite{Green:2010ww} or some other mechanism which can screen the 
scalar mass contributions (see the latter reference for a more complete review).
The broad relation of the underlying physics to the values of $\L_G$ and $\L_S$
is shown in Figure.~\ref{phenoland}. It is striking that the phenomenology of GGM probes the vacuum structure so directly. 
We also show for later reference the exclusions from various phenomenological constraints discussed in detail in Ref.~\cite{Abel:2009ve} for a messenger scale of $10^{10}$\gev.
GGM allows also different
$\Lambda_G^{(i)}$ and $\Lambda_S^{(j)}$ for the different species of gauginos and sfermions although
certain sum-rules still apply~\cite{Meade:2008wd}. However the general parameter space
is prohibitively large for an exhaustive survey and moreover most perturbative models
(for example the direct mediation models, or the hybrid models of \cite{Abel:2009ze})
do correspond to only to single $\L_G$, $\L_S$ and $M_{mess}$ scales. This is 
especially true if one wishes to maintain gauge coupling unification, which is most easily
achieved by keeping an $SU(5)$ structure for the mediating sector.  
In this sense the set of models defined by single $\L_G$, $\L_S$ and $M_{mess}$ scales
are the gauge mediation equivalent of the canonical mSUGRA\footnote{We use the more common term minimal Supergravity (mSUGRA); Constrained MSSM (CMSSM) would be more accurate.}
scenario, with $\L_G$ and $\L_S$ playing the role of the parameters $m_{1/2}$ and
$m_0$ in those {models\footnote{Note that our approach is orthogonal to that taken in Ref.~\cite{Carpenter:2008he} which has
$\Lambda^i_G=\Lambda^i_S=\Lambda^i$, but a different $\Lambda^i$ for each gauge group.}.}

The LHC is currently operating at 7~TeV centre-of-mass energy and, it is hoped, will collect 1fb$^{-1}$ of data by the end of 2011.
It is thus relevant to ask what models and regions of parameter space might be discovered in the next year.
Recent work on this subject includes~\cite{Altunkaynak:2010we,Baer:2010tk,Alves:2010za} and has focussed on the mSUGRA scenario. The goal of this paper is to investigate the signatures and the discovery potential of pure GGM models
at the early LHC stage focussing on collisions at  7~TeV.
In section~\ref{sec:benchmarks} we shall analyse the available parameter space relevant for this regime.
We will proceed to construct a pair of benchmark points with relatively light gluinos. For these we compute the total
$2 \to 2$ production cross-sections, the low-energy spectrum of superpartners and the branching ratios. The NLSP
particles in this region of the parameter space are neutralinos.

We continue in section~\ref{sec:NLSP} with a more general survey of the NLSP phenomenology which is also very relevant
for early stage LHC searches, and analyse other regions of the pure GGM parameter space, complementary to that of
section~\ref{sec:benchmarks}. These will include benchmark points in the stau and co-NLSP regions, and 
a benchmark point in the $\L_G\gg \L_S$ region.

\section{Benchmark points with light gluinos for early LHC discovery}
\label{sec:benchmarks}

{We begin our investigation of the discovery potential
of these models at the early LHC stage by focussing on two explicit benchmark points.
The parameter space of pure GGM models was first investigated in our earlier work \cite{Abel:2009ve}
which also excluded regions due to various constraints. These are shown for the example of 
a $10^{10}$\gev messenger scale in Fig.~\ref{phenoland}.
We will be exploring the allowed regions
of parameter space where either gluino or squark masses are likely to be 
sufficiently light to be discovered with a centre of mass energy of 
up to 7~TeV and integrated luminosity of order $1$\,fb$^{-1}$. As a guideline note that in mSUGRA, 
Ref.~\cite{Baer:2010tk} has argued that when $m_{\tilde{g}} \sim m_{\tilde{q}}$ that the $1$\,fb$^{-1}$ 
reach is approximately 1.1~\tev. Our first two benchmark points will be chosen to have a slightly 
split spectrum with $m_{\tilde{q}} \sim 2-4\, m_{\tilde{g}}$ to allow lighter gluinos.}

\begin{figure}
\begin{center}
\vspace*{-0.6cm}
\subfigure[]{
\includegraphics[bb= 142 75 500 400,clip,width=6.5cm]{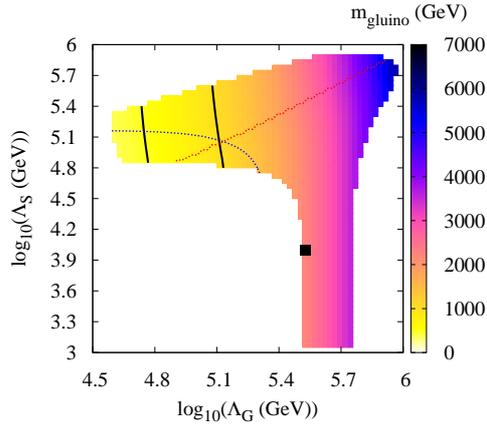}
}
\end{center}
\vspace*{-1.2cm}
\begin{center}
\subfigure[]{
\includegraphics[bb= 142 75 500 400,clip,width=6.5cm]{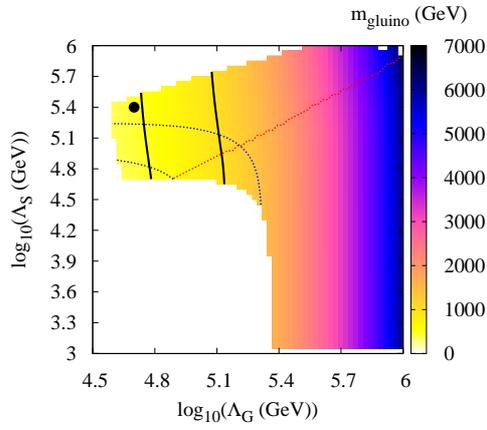}
}
\end{center}
\vspace*{-1.2cm}
\begin{center}
\subfigure[]{
\includegraphics[bb= 142 75 500 400,clip,width=6.5cm]{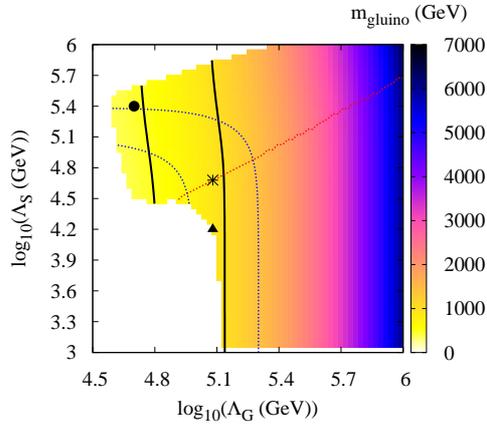}
}
\end{center}
\begin{center}
\caption{
The $\Lambda_G$,~$\Lambda_S$ parameter space for $M_{mess}=10^{8}\gev$ (upper panel), $M_{mess}=10^{10}\gev$ (middle panel) and $M_{mess}=10^{14}\gev$ (lower panel).
Stop mass contours (500\gev and 1\tev) are indicated as dotted lines, and the 500\gev and 1\tev  gluino lines are solid.
The NLSP is neutralino above the dotted red line and stau below. The marked points are the benchmark points discussed in the text:
circular for neutralino NLSP (PGM1a middle panel, PGM1b bottom panel), triangular for stau NLSP (PGM2), a star for stau-neutralino co-NLSP (PGM3)
on the bottom panel and finally a square for PGM4 which has stau NLSP and slepton NNLSP.}
\label{nessie}
\end{center}
\end{figure}

\begin{figure}
\begin{center}
\subfigure[]{
\begin{picture}(190,180)
\includegraphics[bb= 142 80 510 410,width=6.5cm]{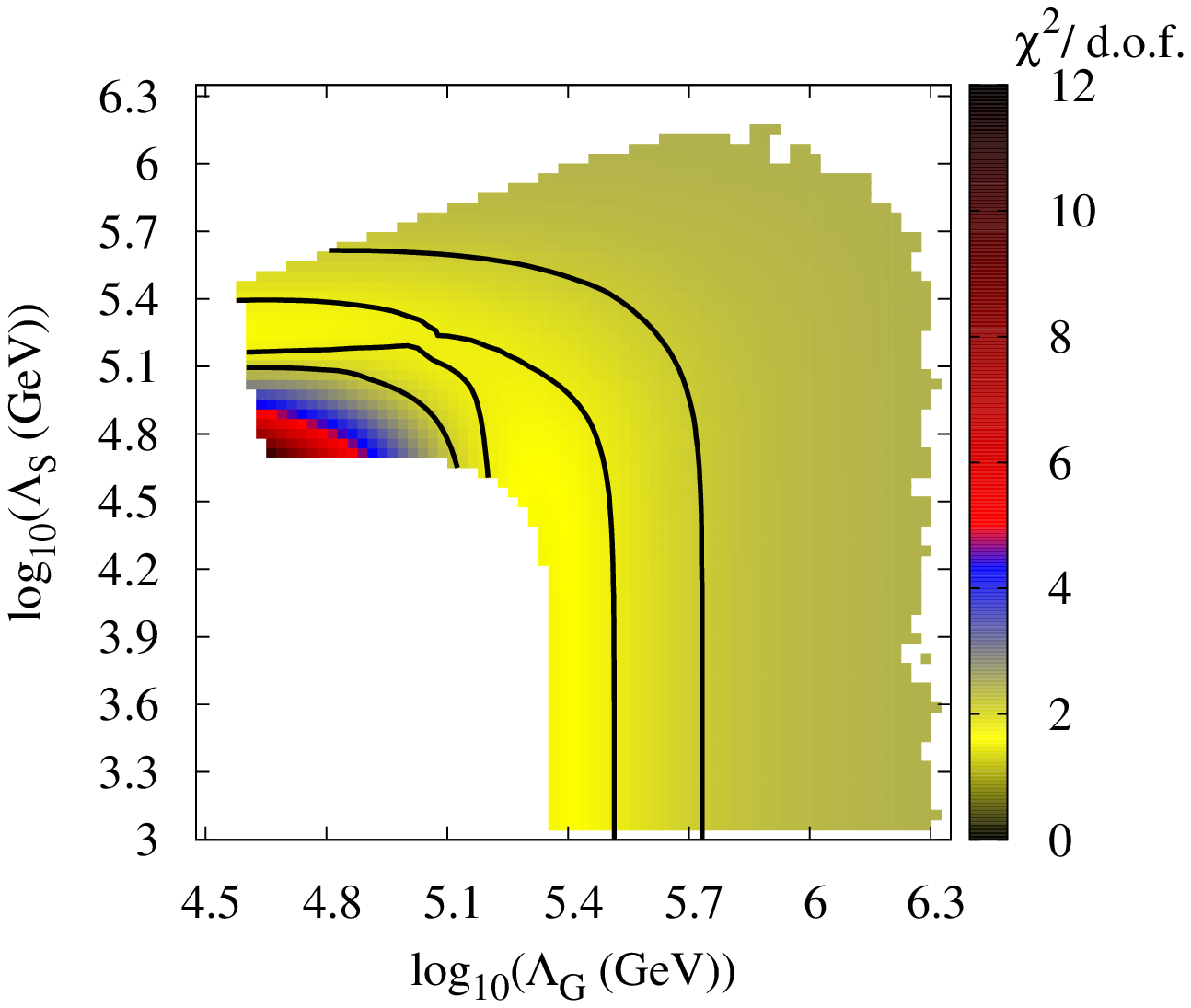}
\Vertex(-140.5,109){2}
\end{picture}
}
\hspace*{1cm}
\subfigure[]{
\begin{picture}(190,180)
\includegraphics[bb= 142 80 510 410,width=6.5cm]{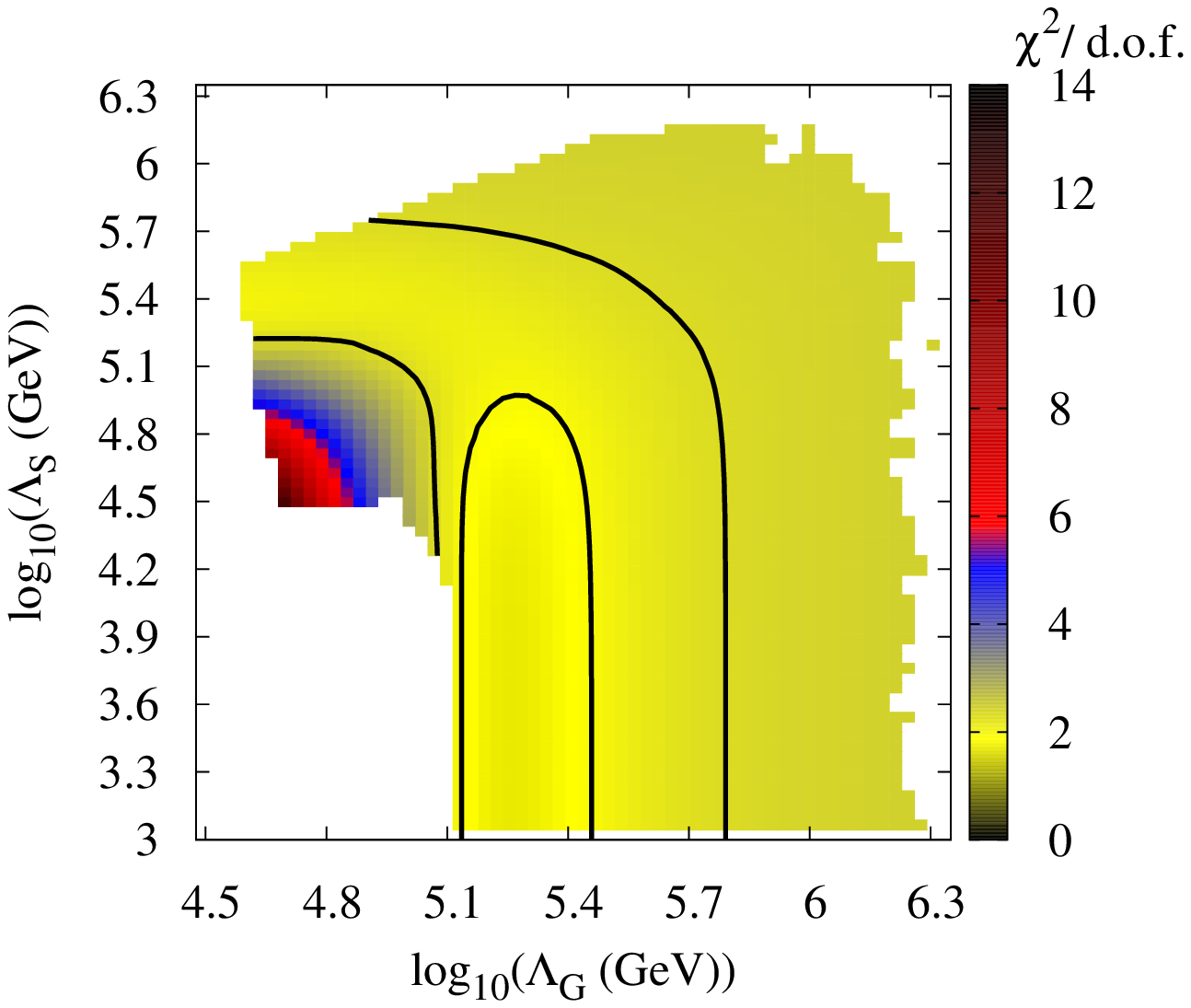}
\Vertex(-140.5,109){2}
\Text(-116,84)[c]{\scalebox{1}[1]{$\star$}}
\Text(-116,66)[c]{\scalebox{1}[1]{$\blacktriangleup$}}
\end{picture}
}
\end{center}
\vspace*{-1cm}
\begin{center}
\caption{Figures (a,b) show the $\chi^2_{tot}$ distribution in the $\L_G$-$\L_S$ plane for $M_{mess}=10^{10}$
and $10^{14}$~GeV respectively. The black lines denote the 68\% and 95\% confidence regions, and we also
show the benchmark points following the same notation as before. The benchmark points are all inside the 95\% confidence regions.}
\label{fig:fits}
\end{center}
\end{figure}

Three scans of the parameter space of pure GGM are shown in Figure~\ref{nessie}, one at $M_{mess}=10^{8}\gev$ one at $M_{mess}=10^{10}\gev$   and one at $M_{mess}=10^{14}\gev$\footnote{It should be noted that lower values of 
messenger scales restrict the parameter space significantly because of our assumption that $B_\mu$ is generated 
radiatively, and the fact that low messenger scales reduce the range of RG running.}. 
In each figure stop mass contours of 500\gev and 1\tev are indicated as dotted lines, and the 500\gev and 1\tev  gluino contours are indicated
as solid lines\footnote{In the $M_{mess}= 10^8$\gev scenario the single dotted contour is for 1\tev stop masses.}. Furthermore, the diagonal dotted red line corresponds to the boundary between neutralino and slepton NLSP. (Note that this 
line is similar to but distinct from the ordinary gauge mediation line of Figure~\ref{phenoland}.)
The figures are also marked with a variety of benchmark points. The circular blobs are benchmark points with a neutralino NLSP,
the triangular points have a stau NLSP, and the stau-neutralino co-NLSP point is indicated by a star. The square blob corresponds to a gaugino mediated point with stau NLSP and slepton NNLSP.
{As will be seen from the $\chi^2$-analysis in Fig.~\ref{fig:fits}, the region
where the squark masses are below 500~GeV is somewhat disfavored by already existing data. Therefore, in this section
we will concentrate on the region of the parameter space with light gluinos -- benchmark points PGM1a and PGM1b.
The triangular, square and star-shaped points with stau NLSP and stau-neutralino co-NLSP will be discussed in Section~\ref{sec:PGM23}.

We chose PGM1a and PGM1b in the light gluino region, to the left of the 500\gev\ line,
marked as circular blobs in each figure.}
The first point (PGM1a) is for a medium to low messenger mass of $10^{10}$~\gev (middle panel of Fig.~\ref{nessie}) .
The second (PGM1b) is for a high messenger mass $10^{14}$~\gev (lower panel of Fig.~\ref{nessie}).
These are typical light gluino points, and as we have said correspond to phenomenology of the ``mildly split'' variety (in which the low energy spectrum is
the Standard Model with only fermionic superpartners)  found in the direct gauge mediation models
analysed in Refs~\cite{Abel:2007jx,Abel:2007nr,Abel:2008gv}.
To some degree these points are quite generic: we chose them to be to the left of the
500\gev gluino line but we have not tried to optimize for the production cross section.
The benchmark points are located in the regions of parameter space which are in good agreement
with currently known experimental constraints.  The experimental constraints were discussed in detail
in Ref.~\cite{Abel:2009ve}. They are quantified by a total $\chi^2$ which it is pleasing to note
is indeed low in these regions.
As shown in Figure~\ref{fig:fits}, where the 68\% and 95\% confidence regions are indicated as black lines, both benchmark points lie well
within the 95\% confidence region.

The spectra of the two benchmark points are given in Table~\ref{tab:sp}, and the neighbourhood of the chosen benchmark points
leads to similar spectra.
The main features of the spectrum in these points are that they have light gluinos with masses below $500$~\gev
and that the NLSP
is a bino-like neutralino\footnote{The LSP is the gravitino as is standard in gauge mediation.}.
Detailed discussion of other possibilities for  NLSP phenomenology in the early stages of the LHC in PGGM
will be presented in section~\ref{sec:NLSP}.

For the rest of the spectrum in Table~\ref{tab:sp} we note that
the first two neutralinos are light, while the Higgsino-like third and fourth neutralinos are much heavier, at the TeV scale.
A similar story holds for the charginos: one is quite light, approximately 135~GeV and is wino-like while the other is
higgsino-like and at the TeV scale. The left-handed sleptons are at the TeV scale, while the right-handed ones vary
from 400-700 GeV depending on the point and sparticle type. As usual, the right-handed staus are the lightest
of the sleptons due to mixing proportional to $\tan\beta$ and the relatively large size of $\lambda_{\tau}$.
Finally, the squarks all have masses above 1~TeV.
Thus for these benchmark points the dominant production channel at the LHC is gluino pair production.

\begin{table}
\begin{center}
\begin{tabular}{|c|c|c|}
\hline
Benchmark point& PGM1a& PGM1b\\\hline
\hline
  $M_{mess}$~(GeV)& $10^{10}$ & $10^{14}$  \\ \hline\hline
$\L_G$~(GeV) & $5\times 10^4$ & $5\times 10^4$  \\ \hline
$\L_S$~(GeV) & $2.5\times 10^5$ & $2.5\times 10^5$  \\ \hline
$\tan\beta$ & 46.6 & 41.2 \\ \hline\hline
$\chi_1^0$ & {\bf 67} & {\bf 67}               \\ \hline
$\chi_2^0$ & 136 & 133  \\ \hline
$\chi_3^0$ & 1038 & 936 \\ \hline
$\chi_4^0$ & 1039 & 938  \\ \hline
$\chi_1^{\pm}$ &  136 & 134 \\ \hline
$\chi_2^{\pm}$ & 1039 & 937\\ \hline
$\tilde{g}$ & {\bf 458} & {\bf 453}      \\ \hline\hline
$\tilde{e}_L,\tilde{\mu}_L$ & 927 & 1013 \\ \hline
$\tilde{e}_R,\tilde{\mu}_R$ & 540 & 712\\ \hline
$\tilde{\tau}_1$ & 392 & 544  \\ \hline
$\tilde{\tau}_2$ & 898 & 964 \\ \hline
$\tilde{\nu}_{1,2}$ & 925 & 1011  \\ \hline
$\tilde{\nu}_3$ & 889 & 958 \\ \hline\hline
$\tilde{t}_1$ & 1418 & 1050\\ \hline
$\tilde{t}_2$ & 1729 & 1471  \\ \hline
$\tilde{b}_1$ & 1578 & 1287 \\ \hline
$\tilde{b}_2$ & 1731 & 1471\\ \hline
$\tilde{u}_L,\tilde{c}_L$ & 2011 & 1760 \\ \hline
$\tilde{u}_R,\tilde{c}_R$ & 1803 & 1520 \\ \hline
$\tilde{d}_L,\tilde{s}_L$ & 1983 & 1734 \\ \hline
$\tilde{d}_R,\tilde{s}_R$ & 1774 & 1460\\ \hline\hline
$h_0$ & 116.9 & 115.3 \\ \hline
$A_0, H_0$ & 944 & 1032  \\ \hline
$H^{\pm}$ & 947 & 1035 \\ \hline\hline
\end{tabular}
\end{center}
\begin{center}
\caption{Spectra for the two benchmark points with light gluinos. All masses are in GeV. The NLSP and the lightest
coloured super-particle (gluino) are shown in bold in each case. These spectra and all other relevant details
can be obtained in SLHA format at \href{http://www.ippp.dur.ac.uk/~SUSY}{\bf http://www.ippp.dur.ac.uk/$\sim$SUSY}}
\label{tab:sp}
\end{center}
\end{table}

We have computed the total production cross-sections to NLO using  PROSPINO~\cite{Beenakker:1996ed,prospino}.
The total gluino production cross sections in pp collissions at 7~TeV are,
\begin{eqnarray}
{\rm PGM1a}:\quad\quad \sigma_{pp\rightarrow\tilde{g}\tilde{g}}=4.09\,{\rm pb}\quad\quad@7\,{\rm TeV}\\\nonumber
{\rm PGM1b}:\quad\quad \sigma_{pp\rightarrow\tilde{g}\tilde{g}}=4.34\,{\rm pb}\quad\quad@7\,{\rm TeV}
\end{eqnarray}
We present cross-sections in femtobarns for various channels in Table~\ref{tab:xsections}.
Since before shutdown the
early-stage
LHC is expected to accumulate approximately 1~fb$^{-1}$ of luminosity,
the entries in the table also give the number of SUSY events expected before then.
The largest contribution to the total production cross-section comes from gluino production, as the gluinos are both relatively light and strongly interacting.
Since for our benchmark points the sfermions are significantly heavier than the gauginos,
production processes involving the squarks are suppressed relative to those only involving gluinos.

Weak gaugino pair production also makes a large contribution to the total cross-section.
Since $\chi_2^0$ is wino-like, the cross-sections for $\chi_2^0 \chi_1^{\pm}$ production are much higher than for the same process with $\chi_2^0$ replaced with the bino-like $\chi_1^0$. Di-chargino production, with a cross-section of 1.32 (1.39)~pb for PGM1a (PGM1b, respectively) also makes an important contribution. All other cross-sections are nearly two orders of magnitude smaller than these, such as $pp\to \tilde{g} \tilde{q}$, also shown in Table~\ref{tab:xsections}.
We have also investigated all the other possibilities, $pp\rightarrow \chi_i^0 \chi_j^{\pm}$, in this family of processes. The Higgsino nature of $\chi_{3,4}^0$ and $\chi_2^{\pm}$ means that production of these particles is negligible. Even though the lightest neutralino has $m_{\chi_1^0}=67$\gev, it is not directly produced in any great numbers.
Of course, these features will change in regions where the hierarchy between
the sfermions and gauginos is less pronounced, and also when the centre of mass energy is raised from 7 to $14$~TeV.
The decays of the lightest chargino are dominated by $\chi_1^{+} \to \chi_1^0 q_{u} \bar{q}_{d}$, which occurs 69\% (70\%) of the time. The rest of the branching ratio is taken up by $\chi_1^{+} \to l^+ \nu_l$, where $l= (e,\mu,\tau)$, with the tau-component taking a somewhat larger share 
of 19\% (20\%).
The wino-like neutralino $\chi^{0}_{2}$ decays predominantly to $\chi_1^0 q \bar{q}$, 71\% (87\%), and to $\chi_1^0 \tau^+ \tau^-$, 23\% (5\%),
with the remaining channels being a combination of $\chi_1^0 l_{e,\mu}^+ l_{e,\mu}^-$ and $\chi_1^0 \nu \bar{\nu}$.
The upshot of this analysis is that in both $\chi_1^{+}\chi_1^{-}$ and $\chi^{\pm}_1\chi_1^0$ production the standard $4j$+ MET analysis should be useful for probing supersymmetry this year.

One might wonder about the existing strong constraints on NLSP neutralino and chargino masses from Tevatron (see \cite{Meade:2009qv} for an overview). The strongest constraints resulting in a high lower bound on neutralino and chargino masses originate from two potential signals. The first one is a di-photon signature studied most recently in Refs.~\cite{Aaltonen:2009tp,Abazov:2010us}. In this case one considers production and subsequent decay
$p\bar{p}\rightarrow \chi^{+}_{1}\chi^{-}_{1}\rightarrow 2\chi^{0}_{1} +\ldots\rightarrow2\gamma+2\tilde{G}+\ldots$ or
$p\bar{p}\rightarrow \chi^{0}_{2}\chi^{\pm}_{1}\rightarrow 2\chi^{0}_{1} +\ldots\rightarrow2\gamma+2\tilde{G}+\ldots$.
The (unobserved) signal is two photons plus missing transverse energy.
However, for such bounds to hold the last decay stage of a neutralino NLSP into a photon and a gravitino must happen promptly (at the very least inside the detector). In general prompt NLSP decays occur only for sufficiently low messenger masses. As we will see in more detail in Sect.~\ref{sec:NLSP}
(see Fig.~\ref{fig:nessie-decaylength}) the NLSP decays happen way outside the detector for our benchmark points PGM1a,b.

\begin{figure}
\begin{center}
  \includegraphics[bb= 182 0 560 600,width=3.2cm]{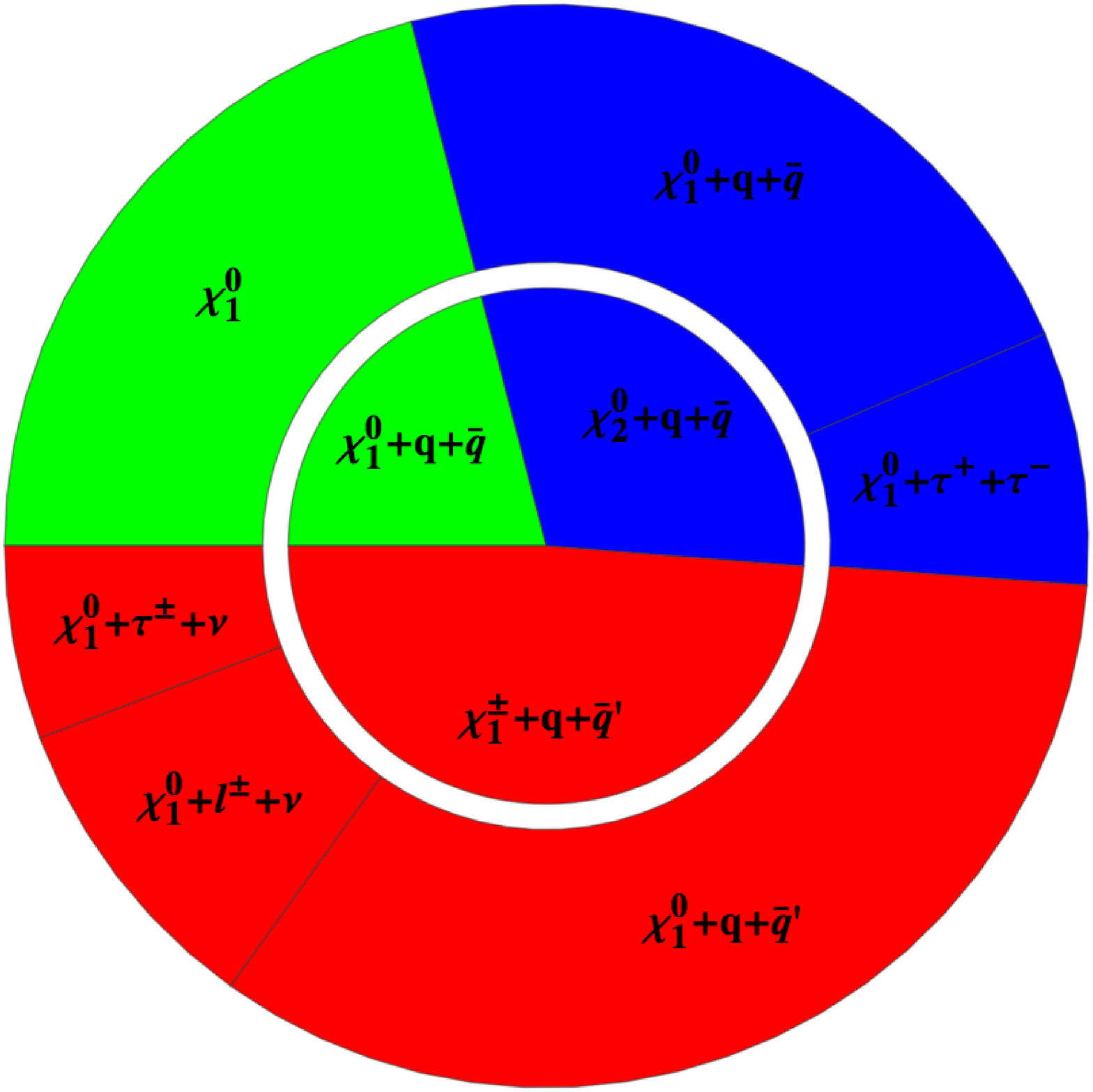}
\hspace*{5cm}
\includegraphics[bb= 182 0 560 600,width=4.0cm]{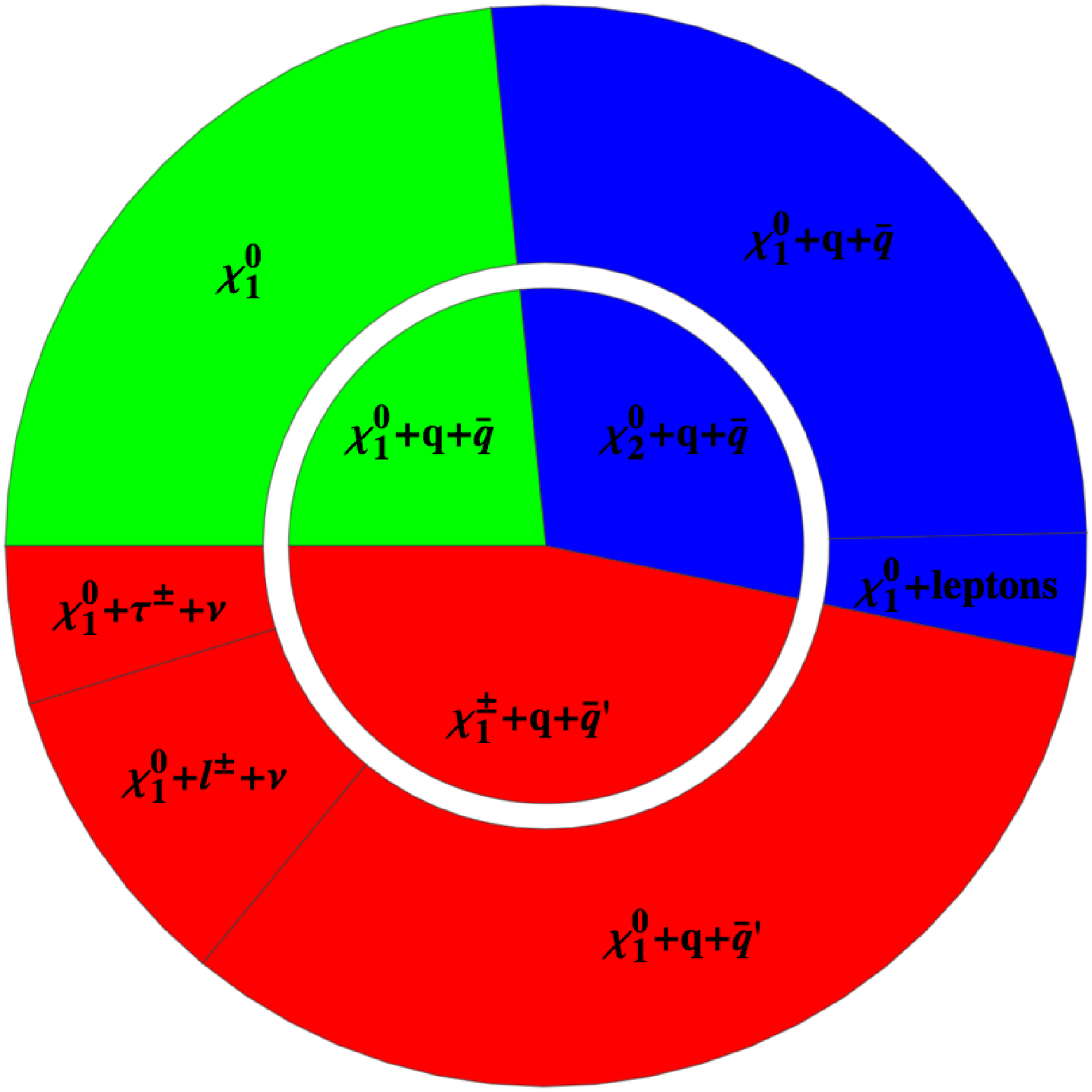}
\vspace*{-0.4cm}
\end{center}
\begin{center}
\caption{Piecharts giving a rough impression of the gluino decay chains/branching ratios
{with the PGM1a benchmark point on the left panel and PGM1b on the right.}
In the first step the gluino decays into the products depicted in the inner ring,
in the next step the daughter sparticle decays into the products given in the outer ring (for simplicity we only write down the additional
decay products for this last decay). We do not display those chains with a branching ratio less than 5\%.}
\label{fig:pies}
\end{center}
\end{figure}

The second signature analysed at Tevatron is a tri-lepton signal. The production would follow from
$p\bar{p}\rightarrow \chi^{0}_{2}\chi^{\pm}_{1}\rightarrow 2\chi^{0}_{1} +\ell \bar{\ell}+\ell^{\prime}\nu$. This signal has been analysed 
in Refs.~\cite{Abazov:2009zi,Forrest:2009gm}  in the context of
mSugra with low values of $\tan(\beta)=3$, setting a new lower limit on chargino masses of 164~GeV.
However, the value of the upper limit depends quite strongly on the choice made for $\tan(\beta)$ as well as other model dependent considerations.
Therefore it will be different in gauge mediation. In particular all our predictions obtained in a pure GGM setup always have much higher values
of $\tan(\beta)$. This increases the branching fraction to $\tau$s which are more difficult to reconstruct. Overall the branching ratios to leptons are quite small in our scenarios as can be seen from the red and blue segments of the outer circles in Fig.~\ref{fig:pies}. This makes the current constraints
inconclusive for the pure GGM predictions analyzed here. This, of course, can be changed by an analysis of (existing) larger sets of Tevatron data, which would be very interesting.

\begin{table}
\begin{center}
\begin{tabular}{|c|c|c|c|c|} \hline
Benchmark Point & $\sigma_{pp\rightarrow\tilde{g}\tilde{g}}$ &  $\sigma_{pp\rightarrow \chi_2^0 \chi_1^{\pm}}$ & $\sigma_{pp\rightarrow \chi_1^{+} \chi_1^{-}}$  & $\sigma_{pp\rightarrow\tilde{g}\tilde{q}}$ \\ \hline
PGM1a & 4090  & 2682 & 1320 & 18.9     \\ \hline
PGM1b & 4340  & 2835 & 1390 & 58.7     \\ \hline
\end{tabular}
\vspace{0.3cm}
\caption{Production rates for the most important processes for the two
benchmark points under consideration at the LHC with $\sqrt{s}=7$~TeV. All cross-sections are in femtobarns.}
\label{tab:xsections}
\end{center}
\end{table}

\noindent We now focus on $pp\rightarrow \tilde{g}\tilde{g}$, and discuss the main decay avenues to the final states including NLSPs.
It can be seen from this analysis that the gluino decays dominantly into a chargino plus a quark and an antiquark.
Subsequently the chargino decays into a neutralino plus either a quark and an antiquark, or a lepton and a neutrino, as discussed above.
An alternative interesting channel is that each gluino decays directly into a neutralino and a quark-antiquark pair.
In all of these processes the two gluinos will decay
into a total of 4 or more coloured particles and two neutralinos (plus leptons in some cases).
In Fig.~\ref{fig:pies} the branching ratios of the gluino
{and the daughter sparticle decays
are represented graphically, with the PGM1a benchmark point shown on the left panel and the PGM1b point on the right.
Decay chains with branching ratios of less than 5\% are not
shown\footnote{
For example, for the PGM1a the  $\chi_2^0$ decays to 23\% into $\tau$'s but only to less than 4\% into other leptons
and thus the latter are not shown on the left panel in Fig.~\ref{fig:pies}.
For the PGM1b the $\chi_2^0$ decays  5\% into $\tau$'s and to nearly 4\% into other leptons, these two contributions are combined
and collectively called ``leptons'' on the right panel in Fig.~\ref{fig:pies}.}.
}

The full set of branching ratios (as well as the spectra in SLHA format) for these benchmark points can be found at

{\vspace{-0.5cm} \centering{
\href{http://www.ippp.dur.ac.uk/~SUSY}{\bf http://www.ippp.dur.ac.uk/$\sim$SUSY}
}\\\vspace{0.4cm}}

In the following section we will present a more general overview of the NLSP phenomenology.
We shall then perform a complementary analysis, in regions of the parameter space where the NLSP is a stau or a
light slepton or there are co-NLSPs (in practice these are areas where the stau and neutralino are nearly degenerate in mass).
Again we focus on areas that may be relevant to the early LHC searches.

\section{Survey of NLSP phenomenology}
\label{sec:NLSP}

In gauge mediated models the Lightest Supersymmetric Particle (LSP) is always the gravitino \cite{Giudice:1998bp}. There is
much interest therefore in the phenomenology of the {\em Next-to}-LSP (NLSP)  as this is the metastable state into which any produced
superpartner will decay before ultimately decaying to the gravitino. Therefore
it is instructive to map out the NLSP phenomenology in the whole $\L_G$, $\L_S$ parameter space,
and describe in more detail some of the top-down models that correspond to the different regions.

For the assumptions we
outlined above, the NLSP is either slepton or neutralino. The NLSP phenomenology is of great
interest for two reasons~\cite{Giudice:1998bp}. First it is typically very long lived -- its decay to the gravitino is suppressed:
$\Gamma \propto \, m_{NLSP}^5/F_0^2$ where $m_{NLSP}$ is its mass and $F_0$ is the intrinsic scale of supersymmetry breaking
in the hidden sector (i.e. the potential is $\langle V\rangle =F_0^2$).
Typically, depending on how the SUSY breaking encoded by $F_0$ is mediated, $\Gamma$ represents many orders of magnitude of suppression.
If it is
sufficiently long lived the NLSP will exit the detector as missing energy, or {leave a muon-like track} if it is charged (e.g. if it is
a stau).  On the other hand for certain values of parameters (which we discuss presently) the particle can decay inside
the detector possibly allowing one to resolve a displaced decay vertex.
Moreover such a measurement would give direct information about the SUSY breaking in the hidden sector $F_0$ rather than
that seen in the visible sector which depends heavily on the particular type of (gauge) mediation.

\begin{figure}
\begin{center}
\vspace*{-0.6cm}
\subfigure[]{
\includegraphics[bb= 142 75 500 400,clip,width=6.5cm]{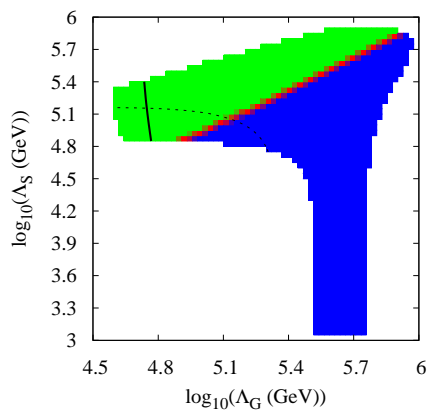}
}
\end{center}
\vspace*{-1.2cm}
\begin{center}
\subfigure[]{
\includegraphics[bb= 142 75 500 400,clip,width=6.5cm]{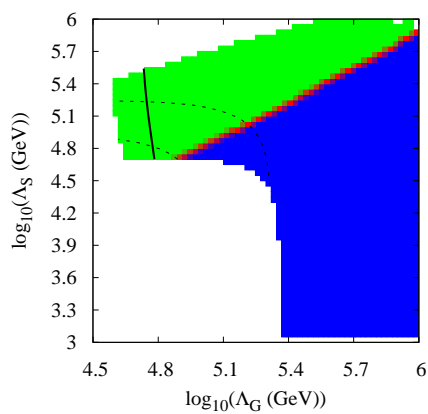}
}
\end{center}
\vspace*{-1.2cm}
\begin{center}
\subfigure[]{
\includegraphics[bb= 142 75 500 400,clip,width=6.5cm]{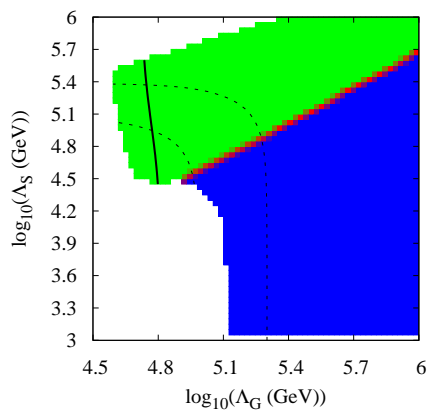}
}
\end{center}
\begin{center}
\caption{The NLSP regions in the $\Lambda_G$,~$\Lambda_S$ parameter space for $M_{mess}=10^{8}\gev$ (top figure), $M_{mess}=10^{10}\gev$ (middle figure)
and $M_{mess}=10^{14}\gev$ (bottom figure).
The NLSP is $\chi_1^0$ in the green region, $\chi_1^0 /\tilde{\tau}$ co-NLSP in the red region and
$\tilde{\tau}$ in the blue region.}
\label{fig:nessie-nlsp}
\end{center}
\end{figure}
The compositions of the NLSP in different regions of parameter space are shown in Figure~\ref{fig:nessie-nlsp}, again
one at $M_{mess}=10^{8}\gev$ (top), one at $M_{mess}=10^{10}\gev$ (middle) and one at $M_{mess}=10^{14}\gev$ (bottom).
In each figure again stop mass contours of 500\gev and 1\tev are indicated as dotted lines, and the 500\gev  gluino contour is indicated
as a solid line.
 We have indicated 3 different NLSP regions on the figures,
each giving quite distinct experimental signatures:

\begin{itemize}
\item Neutralino NLSP (Marked in green): no ionization track and either missing energy or displaced vertex with decay predominantly to photon ($\chi^0_1 \rightarrow \tilde{G} \gamma$) or jet/lepton pairs ($ \chi^0_1 \rightarrow \tilde{G} Z\rightarrow \tilde{G} +jets/l{\bar l}$).
\item Stau NLSP (Marked in blue): ionization track plus possible displaced vertex with decay predominantly to jets ($ \tilde{\tau}_R \rightarrow \tilde{G} \tau\rightarrow \tilde{G}\nu_\tau +jets/l'{\bar l}$).

\item Neutralino/stau co-NLSP (Marked in red):
if the mass difference between the neutralino and stau is less than $m_\tau$, then the NNLSP is unable to decay to the
NLSP, and each component behaves effectively a separate NLSP.  One expects a mix of those previous two cases.
\end{itemize}

We can treat the decay length of the NLSP as follows. First consider the decays: they go through the interaction
term which for on-shell particles
is~\cite{Giudice:1998bp}
\be
\label{int}
{\cal L} = \frac{1}{F_0} \left( (m_f^2-m_{\tilde{f}}^2)
\bar{f}_L \tilde{f} + \frac{M_{\tilde{\lambda}_i}}{4\sqrt{2}} \bar{\tilde{\lambda}}_i \sigma^{\mu\nu}F^i_{\mu\nu}
\right) \tilde{G} + h.c.
\ee
where $\tilde{G}$ is the Goldstino and as we have already stated $F_0$ is the absolute scale of supersymmetry breaking.
The decay length  derived from Eq.\eqref{int} is given by
\be
\label{f0-eq}
L_{decay} = \frac{1}{\kappa} \left( \frac{100\gev }{m_{NLSP}} \right)^5
\left( \frac{F_0 }{(100\tev)^2 } \right)^2 0.1\, {\rm mm}
\ee
where the factor $\kappa $ is a calculable number depending on the mixing in the NLSP, and is of order unity (precisely unity for the stau in fact).
The interesting case is when decay takes place inside the detector which conservatively requires $L_{decay}< 10$\,m. For  NLSP masses less that
500\gev, this translates into
\be
\label{eq-fbound}
\sqrt{F_0} \lesssim 10^4 \tev \, .
\ee
Thus $F_0$ will be at the lower end of the possible range.

In order to get more precise information we need to consider the relation between $F_0$ and $\Lambda_G$ or $\Lambda_S$. This is very model dependent,
but simplifies if we take there to be only one source of supersymmetry breaking (i.e. one potential Goldstino) and one dominant source of mediation
for gauginos or scalars. Under this assumption
the relation between the $\Lambda$'s and $F_0$ can be expressed with two parameters $k_G$ and $k_S$ as
\be \label{eqn:10}
\Lambda_G = k_G F_0/M_{mess} \,\, ; \, \,
\Lambda_S = k_S F_0/M_{mess} \, .
\ee
In GGM, $k_G$ and $k_S$ are independent parameters which encode the difference between the gauge and scalar mass scales $\L_G$ and $\L_S$.
In ordinary gauge mediation, $k_G=k_S,$ and this corresponds to a simple one-scale special case of GGM.
In general, as will be reviewed shortly, the range of values for $k_G$ and $k_S$ is highly model-dependent.

In order to present model-independent information
it is useful to express $F_0$ with reference to $\Lambda_G$: i.e. we replace $F_0 = k_G^{-1}\L_G M_{mess}$.
The decay length $L_{decay}$ derived from Eq.\eqref{int} is given by
\be
\label{eq-withk}
{k_G^{2}} L_{decay} = \frac{1}{\kappa} \left( \frac{100\gev }{m_{NLSP}} \right)^5
\left( \frac{\sqrt{\Lambda_G M_{mess}} }{100\tev } \right)^4 0.1\, {\rm mm}
\ee
We then plot contours of ${k_G^{2}} L$. The reason that this is a most useful parameterization is that
in the regions where $\Lambda_G>\Lambda_S$ the NLSP is mainly slepton,
as can be seen from Fig.~\ref{fig:nessie-nlsp},
and its mass is dominated by
renormalization group contributions from the gauginos (except when $\Lambda_G/\Lambda_S \sim \mathcal{O}(1-10)$).
Thus $m_{NLSP}$ is mainly a function of $\L_G$ (just as the stop mass is in fact).
On the other hand in the regions where $\Lambda_G<\Lambda_S$ the NLSP is mainly a bino-like
neutralino and again its mass is
expected to be dominated by $\L_G$. Hence the RHS of Eq.~\eqref{eq-withk} is predominantly a function of $\Lambda_G$.

\begin{figure}
\begin{center}
\vspace*{-0.6cm}
\subfigure[]{
\includegraphics[bb= 142 75 500 400,clip,width=6.5cm]{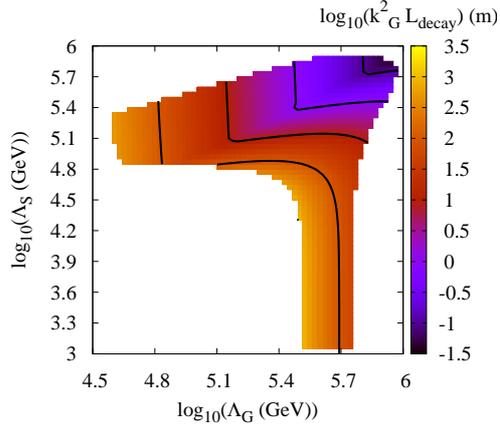}
}
\end{center}
\vspace*{-1.2cm}
\begin{center}
\subfigure[]{
\includegraphics[bb= 142 75 500 400,clip,width=6.5cm]{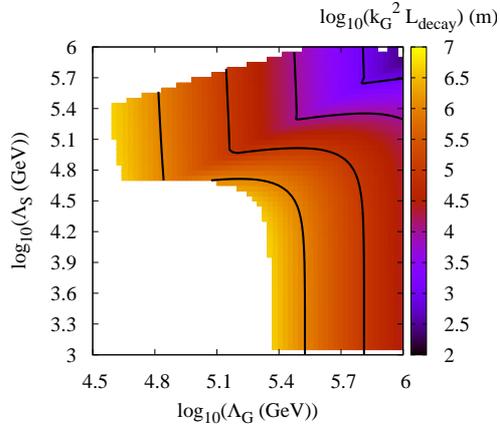}
}
\end{center}
\vspace*{-1.2cm}
\begin{center}
\subfigure[]{
\includegraphics[bb= 142 75 500 400,clip,width=6.5cm]{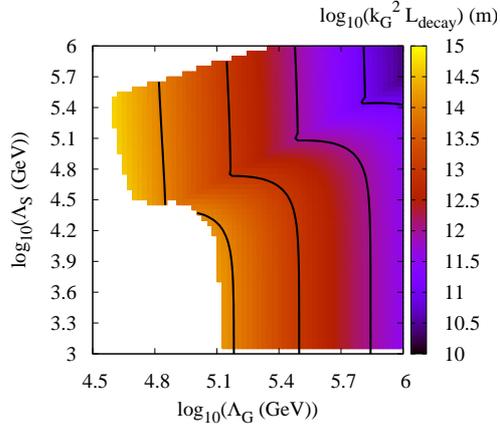}
}
\end{center}
\begin{center}
\caption{
This figure shows the logarithm of the decay length in meters of the NLSP, $\log_{10}(k_G^2 L_{decay})$
for $M_{mess}= 1\times 10^{8}$~GeV (top), $M_{mess}= 1\times 10^{10}$~GeV (middle) and $M_{mess}= 1\times 10^{14}$~GeV (bottom), as well as contours for each case.}
\label{fig:nessie-decaylength}
\end{center}
\end{figure}

We show the results for the decay lengths $\log_{10}(k_G^2 L_{decay})$ in Figure~\ref{fig:nessie-decaylength} for the three values of the messenger mass.
We see that the contours follow a vertical, horizontal and vertical again pattern, which we now explain.
Starting at the top of the figures, when $\Lambda_S$ is large the NLSP is the neutralino, and the decay
length does not change with decreasing $\Lambda_S$ as both $m_{NLSP}$ and $\Lambda_G$ are constant.
When the NLSP species changes from neutralino to the lightest stau, there is a kink in the contour.
This is partly due to the change in $\kappa$, and also to the change in the behaviour of the NLSP mass
with $\Lambda_G$ and $\Lambda_S$. In this regime the stau mass is dominated by $\Lambda_S$ and,
although $k_G^2 L_{decay}$ is proportional to $\Lambda_G^2$ the factor of $1/m_{\tilde{\tau}}^5$
means that $k_G^2 L_{decay}$ is proportional to $1/\Lambda_S^5$. When these two parameters are of
the same of order of magnitude the contour thus appears flat in $\Lambda_S$.
Finally, when $\Lambda_G / \Lambda_S \sim 10$ the stau mass begins to be dominated by $\Lambda_G$
and generated mostly through RG running and so the contour is again approximated by a line of constant $\Lambda_G$.

It is instructive to now consider the values of $k_G$ that one expects to have in various different
top-down scenarios  in order to see whether decays inside the detector are a possibility:
\begin{itemize}

\item{Ordinary mediation}: Here one has only one messenger and $\Lambda_G=\L_S$ and
$k_G$ is the coupling of the messenger to the SUSY breaking $F$-term. Typically one
takes $k_G\sim 1$. In this case Figure~\ref{fig:nessie-decaylength} gives directly the decay lengths of the NLSP.
Evidently low messenger scales are required for decay inside the detector.
For $M_{mess}=10^8\gev$ decays {can happen inside or outside the detector, depending on the region of parameter space. Comparing Fig.~\ref{fig:nessie-decaylength} with Fig.~\ref{nessie} we see that decay inside the detector happens when $m_{\tilde{g}} \geq 1 $\tev}.
 Intermediate scales $M_{mess}=10^{10}\gev$ would require high
values of $\Lambda_G,\,\L_S$ which leads to very high masses outside the early discovery region.

\item{Suppressed ordinary gauge mediation}: Ref.~\cite{Murayama:2006yf} presented a simple scheme for gauge mediation
in which a single messenger field was coupled to a metastable SUSY-breaking sector of the type introduced in Ref.~\cite{Intriligator:2006dd}.
In these models the Goldstino superfield is a composite particle (a ``meson'') and hence the effective coupling to the messenger fields is
suppressed by a factor $k_G\sim k_S \sim \frac{\Lambda_{comp}}{M_X} \ll 1$ where $M_X$ is some
{high fundamental scale which might be $M_{\rm Pl}$,} and $\Lambda_{comp}$
is the scale of compositeness. The general expectation is that $k_G,~k_S\ll 1$ and indeed phenomenological viability demands it.
For example the values chosen in Ref.~\cite{Murayama:2006yf} give $k_G,~k_S \sim 10^{-7}$. Hence decay inside the detector
(or indeed the Solar system)  is clearly impossible for any values of $M_{mess}$ or $\Lambda_G$,~$\Lambda_S$.

\item{Mildly split spectrum}: phenomenology of the ``mildly split'' variety (in which the low energy spectrum is
the Standard Model plus only the fermionic superpartners)  was found in the direct gauge mediation models
analysed in Refs~\cite{Abel:2007jx,Abel:2007nr,Abel:2008gv}. This type of phenomenology is in fact
characteristic of models that have no {\em tree-level} metastability, due to a theorem by Komargodski
and Shih \cite{Komargodski:2009jf} that tree-level gaugino masses are equivalent to there existing
some point in moduli space where there is a tachyon. These models have $\Lambda_G\ll \L_S\lesssim F_0$
and hence they correspond to $k_S\sim 1$ and $k_G\sim 10^{-(1-2)}$. Since the NLSP mass is governed by $\Lambda_G$,
viable phenomenology requires larger values of $F_0$ and a commensurately slower  NSLP decay. For example low
messenger scales $M_{mess}=10^8\gev $ can just give decay within the detector whereas already intermediate scales
$M_{mess}=10^{10}\gev $ do not allow decay within the detector at all.

\item{Many messenger/strong coupling limit}: In ordinary gauge mediation, the gaugino mass scale $\Lambda_G$ is proportional to the number of messengers $N_{mess}$, and the mediated squark mass
scale $\Lambda_S$ is proportional to $\sqrt{N_{mess}}$. Thus in the ``many effective messengers'' limit,
we access the $\Lambda_G\gg \Lambda_S$ region of the parameter space and moreover we can effectively have $k_G\gg 1$,
so the NLSP decays more rapidly. For example when $k_G\sim 30$ even intermediate messenger masses, $M_{mess}=10^{10}\gev$,  allow NLSP
decays to take  within the detector with reasonably low masses for the coloured sparticles (i.e. below 1TeV).
Of course naively adding many messengers leads to   
strong coupling in the visible sector: as discussed in the introduction calculable models in this region require some mechanism to screen the scalar mass contribution.

\end{itemize}

To summarise the discussion arising from Fig.~\ref{fig:nessie-decaylength}: In most cases NLSP decay happens well outside the detector. Decays inside the detector are only possible for relatively low messenger masses, high SUSY breaking scales and/or quite strong coupling $k_{G}$ to the hidden sector.

\subsection{Stau and co-NLSP Benchmark points}

\label{sec:PGM23}

As can be seen from Figure~\ref{fig:nessie-nlsp}, the stau NLSP and co-NLSP regions both have $m_{\tilde{g}}>500\gev$ and $m_{\tilde{q}}>500\gev$.
The low-mass parts of these regions are also disfavoured according to the analysis of supersymmetric contributions to Standard Model
observables in~\cite{Abel:2009ve}. Accordingly the production cross-sections in these cases are lower than for neutralino NLSP.
However, in the stau NLSP case (and also possibly in the co-NLSP scenario) with higher messenger scales the stau is stable on
collider length- and time-scales. The signatures from such charged massive metastable particles (CHAMPS) are unique enough that
early SUSY discovery may be feasible even with the smaller cross-sections in this scenario.
{We have therefore selected two benchmark points, PGM2 with
a stau NLSP, and PGM3 with a stau co-NLSP (both shown on the bottom panel in Fig.~\ref{nessie})
and performed a preliminary analysis of their phenomenology.} In addition we have for completeness 
 chosen a fifth benchmark point, PGM4, at low messenger scales in the $\L_G\gg\L_S$ region (with $\L_G=3.4\times 10^5$\gev and 
$\L_G= 10^4$\gev). 
As for the light gluino points,
the SLHA files and are available on the Pure GGM website mentioned above. 
{The spectra are shown in Table~\ref{tab:sp2}.}

\begin{table}
\begin{center}
\begin{tabular}{|c|c|c|c|}
\hline
Benchmark point& PGM2& PGM3 & PGM4\\\hline
\hline
  $M_{mess}$~(GeV)& $10^{14}$ & $10^{14}$ & $10^8$  \\ \hline\hline
$\L_G$~(GeV) & $1.2\times10^5$ & $1.2\times10^5$ & $3.4\times10^5$  \\ \hline
$\L_S$~(GeV) & $1.6\times10^4$ & $4.76\times10^4$ & $10^4$  \\ \hline
$\tan\beta$ & 19.0 & 20.5 & 34.4 \\ \hline\hline
$\chi_1^0$ & 156  & {\bf 157} & 456               \\ \hline
$\chi_2^0$ & 292 & 296 & 723 \\ \hline
$\chi_3^0$ & 461 & 489 & 743 \\ \hline
$\chi_4^0$ & 479 & 504 & 897  \\ \hline
$\chi_1^{\pm}$ &  291 & 295 & 720 \\ \hline
$\chi_2^{\pm}$ & 480 & 505 & 898 \\ \hline
$\tilde{g}$ & 879 & 887 &  2239    \\ \hline\hline
$\tilde{e}_L,\tilde{\mu}_L$ & 246 & 305 & 406 \\ \hline
$\tilde{e}_R,\tilde{\mu}_R$ & 129 & 182 & 163\\ \hline
$\tilde{\tau}_1$ & {\bf 100} & {\bf 157}  & {\bf 110} \\ \hline
$\tilde{\tau}_2$ & 254 & 310 & 423\\ \hline
$\tilde{\nu}_{1,2}$ & 234 & 296 & 401 \\ \hline
$\tilde{\nu}_3$ & 232 & 293 & 401 \\ \hline\hline
$\tilde{t}_1$ & 618 &  650 & 1459\\ \hline
$\tilde{t}_2$ & 786 & 823  & 1601\\ \hline
$\tilde{b}_1$ & 726 & 769 & 1557\\ \hline
$\tilde{b}_2$ & 761 & 802 & 1596\\ \hline
$\tilde{u}_L,\tilde{c}_L$ & 804 & 860 & 1682\\ \hline
$\tilde{u}_R,\tilde{c}_R$ & 766 & 810 & 1621\\ \hline
$\tilde{d}_L,\tilde{s}_L$ & 795 & 850 & 1658\\ \hline
$\tilde{d}_R,\tilde{s}_R$ & 765 & 805 & 1621\\ \hline\hline
$h_0$ & 113.3 & 113.4 & 118\\ \hline
$A_0, H_0$ & 493 & 539 & 781 \\ \hline
$H^{\pm}$ & 499 & 545 & 785\\ \hline\hline
\end{tabular}
\end{center}
\begin{center}
\caption{
Spectra for three benchmark points with stau NLSP. PGM2 has slepton NNLSP and a high messenger scale and PGM3 has stau-neutralino co-NLSP also at a high messenger scale. PGM4 is at low messenger scale with slepton NNLSP. All masses are in GeV. The NLSP is shown in bold in each case. These spectra and all other relevant details
can be obtained in SLHA format at \href{http://www.ippp.dur.ac.uk/~SUSY}{\bf http://www.ippp.dur.ac.uk/$\sim$SUSY}}
\label{tab:sp2}
\end{center}
\end{table}

Let us first consider the stau NLSP case, {PGM2}. Due to the constraint from the Higgs mass, it is not possible to have very light squarks in this case.
The point we have chosen has $\Lambda_G=1.2\times10^5$ and $\Lambda_S = 1.6\times10^4$, which corresponds to a moderately large value of $\tan\beta = 19$.
The squark masses for our benchmark point are in the range $750-800\gev$, while the mass of the lightest stop is 617\gev.
The gluino mass is slightly heavier at 880\gev. The lightest stau mass is 100\gev, just above the bound from direct searches,
and the lightest neutralino mass is 156\gev. The stau-smuon splitting is
{29\gev.}
We now turn to the production cross-sections
for this point. As the gluino mass {in PGM2} is nearly double that of the neutralino NLSP {in points PGM1a and PGM1b,}
the $pp\to\tilde{g}\tilde{g}$
cross-section is much smaller. The processes with the largest production cross-sections for the stau NLSP benchmark point
{PGM2} are shown
in Table~\ref{tab:stau-xsections} in femtobarns. While the squark production cross-sections are higher than for the PGM1 scenarios,
for this point the total number of SUSY events will be about 600, when one includes the processes with smaller contributions.
While we have not performed a detailed simulation, the {PGM2} point should just be within the range of discovery of the ATLAS detector
in the first year of operation~\cite{Raklev:2009mg}. In the stau NLSP scenario one does not expect any missing $E_T$ since
the pair produced staus will turn up in the calorimeters at the end of the SUSY cascade.
{From the strong production channels $pp\to\tilde{g}\tilde{g}$ and $pp\to\tilde{g}\tilde{q}$ we expect $\geq 2$ jets plus two muon-like objects.
In addition we also have significant $\tilde{\tau}$ pair production which should just give two muon-like objects.
Together these channels should provide good chances for early SUSY discovery in these scenarios.}
Finally, single production of neutralinos and charginos in conjuction with a gluino or a squark is negligible.

\begin{table}
\begin{center}
\begin{tabular}{|c|c|c|c|c|c|c|} \hline
Benchmark Point & $\sigma_{pp\rightarrow\tilde{g}\tilde{g}}$ & $\sigma_{pp\rightarrow\tilde{q}\tilde{q}}$ &
$\sigma_{pp\rightarrow\tilde{g}\tilde{q}}$ & $\sigma_{pp\rightarrow\tilde{q}\bar{\tilde{q}}}$
& $\sigma_{pp\rightarrow \tilde{\tau}_i \tilde{\tau}_j}$ & $\sigma_{pp\rightarrow \chi_2^0 \chi_1^{\pm}}$  \\ \hline
PGM2 & 17 & 190 & 164 & 54 & 91 & 49           \\ \hline
PGM3 & 16 & 133 & 128 & 34 & 17 & 50           \\ \hline
\end{tabular}
\end{center}
\begin{center}
\caption{This table shows the production rates for the most important processes for the
{stau (PGM2) and co-NLSP (PGM3) benchmark points}
 at the LHC with $\sqrt{s}=7$~TeV. All cross-sections are in femtobarns.}
\label{tab:stau-xsections}
\end{center}
\end{table}

Next, we discuss the possibility of a stau-neutralino co-NLSP.
If we were to decrease $\Lambda_G$ very much, this would lead to an unacceptable decrease in the Higgs mass.
Therefore we must increase $\Lambda_S$ in order to achieve $m_{\tilde{\tau}} \sim m_{\chi_1^0}$. The co-NLSP point {PGM3}
has $\Lambda_G=1.2\times 10^5$, $\Lambda_S = 4.76\times 10^4$ and $\tan\beta=20.5$.
The point we have selected has $m_{\tilde{\tau}_1}=157\gev$ and $m_{\chi_1^0} = 157\gev$, with neutralino marginally heavier than the stau.
As the scalar mass parameter $\Lambda_S$ has increased somewhat, the squark masses are heavier at this point by around 50\gev compared
with the stau NLSP point. The slepton masses are also higher, and the light smuon and selectron masses are 181\gev.
The production cross-sections are broadly similar to the stau NLSP case, but somewhat smaller due to the higher masses
and more compressed spectrum in this case.

Finally we discuss the stau NLSP point in the many messenger limit, PGM4. This point has the interesting feature that the lightest neutralino is heavier than all the sleptons and sneutrinos. The phenomenology of this scenario has been explored in~\cite{DeSimone:2008gm,DeSimone:2009ws}, and includes the presence of many leptons from decay chains leading to the NLSP. It is not possible in PGGM to achieve low enough coloured sparticle masses to have large gluino and squark production cross-sections. The reason for this is as follows. The scalar masses in this gaugino mediated region are generated predominatly by RG running, and take the form 
\begin{equation}
\delta m^2_{\tilde{f}} \sim \frac{\alpha}{4\pi} \Lambda_G^2
\end{equation}
where a summation over the gauge groups is implied. The main constraint on the value of $\Lambda_G$ in the gaugino mediated region are the direct search constraints, and specifically the constraint on the mass of the stau. The staus are only weakly interacting, and thus require relatively large values of $\Lambda_G$ to evade the direct search constraints. This large $\Lambda_G$ is what causes the strongly interacting sparticles to have such large masses. In the full GGM parameter space with three independent gaugino masses one could increase coloured sparticle production by keeping $\Lambda_G^{1,2}$ fixed and decreasing $\Lambda_G^3$. This would leave the slepton, neutralino and chargino masses fixed while decreasing the squark and gluino masses.
Accordingly sparticle production at PGM4 at LHC7 is mostly due to direct production of the stau NLSP. This has a cross-section of 62~fb. Almost all the produced staus are the NLSP however (the cross section into these being 61.8 fb). Thus the 
leptogenic signals due to heavy stau or neutralino decay described in Ref.\cite{DeSimone:2009ws} will not be a feature of 
the LHC at 7 TeV in the pure GGM scenario, and will only appear at higher energies. The main signal in this region for the moment will be an excess of di-muon events, and possibly the displaced vertex signals of NLSP decay inside the detector.

\section{Conclusions}

We have made a survey of the phenomenology of Pure General Gauge Mediation -- i.e. in which the $B_{\mu}$ parameter is generated radiatively, with a particular emphasis on its testability in early LHC searches (at 7~TeV). Five benchmark points were presented: two corresponding to  light gluino regions ($m_{\tilde{g}}\lesssim 500$~GeV with a bino-like neutralino NLSP), two to a stau NLSP and one to stau/neutralino co-NLSP. These benchmark points are representative of the different phenomenology that can occur in the regions of parameter space. We presented a preliminary analysis of the spectrum, production cross sections and branching ratios, which suggests that all of these points can be discovered in the first year of LHC running with appropriate selection cuts.  The full set of data in SLHA format for these benchmark points can be found at

{
\centering{
\href{http://www.ippp.dur.ac.uk/~SUSY}{\bf http://www.ippp.dur.ac.uk/$\sim$SUSY}
}\\\vspace{0.5cm}}

\noindent We also surveyed and discussed NLSP phenomenology in this set-up, focussing on the possibility of
NLSP decays inside the detector in various different schemes of SUSY breaking. Pure GGM with medium to low messenger masses
($10^{6-10}$~GeV) can give detectable decays with displaced vertices
inside the detector, and hence direct knowledge of the fundamental scale of SUSY breaking.

\subsection*{Acknowledgements}

We thank Yuri Gershtein and Zohar Komargodski 
for interesting discussions. MJD thanks St John's College, the CET and EPSRC for financial support.
{SAA and VVK are in receipt of Leverhulme Research Fellowships.} 


\providecommand{\href}[2]{#2}\begingroup\raggedright
\endgroup
\end{document}